\documentclass[twoside]{article}

\usepackage{amssymb,latexsym,amsmath}     
\usepackage[utf8]{inputenc} 
\usepackage[T1]{fontenc}    
\usepackage{hyperref}       
\usepackage{url}            
\usepackage{booktabs}       
\usepackage{amsfonts}       
\usepackage{nicefrac}       
\usepackage{microtype}      
\usepackage{lipsum}
\usepackage{fancyhdr}       
\usepackage{graphicx}       
\usepackage{array}          
\usepackage{tabularx}       
\graphicspath{{media/}}     

\title{Multiscale Correlation of Morphological, Chemical, and Optical Properties in p-Type Porous Silicon}

\author{
  A. Ramírez-Porras$\thanks{Corresponding author: \texttt{arturo.ramirez@ucr.ac.cr}}$,
  I. Prado-Bermúdez\\[1ex]
  \small Escuela de Física and Centro de Investigación en Ciencia e Ingeniería de Materiales (CICIMA),\\
  \small Sede Rodrigo Facio, Universidad de Costa Rica, San Pedro 11501, Costa Rica\\
  }

\begin{document}
\maketitle

\begin{abstract}
We report a systematic study of the correlations among electrochemical processing conditions and the morphological, chemical, and optical properties of porous silicon (pSi). Twenty pSi samples were fabricated from boron-doped (100) crystalline silicon by varying current density, etching time, and hydrofluoric-acid concentration. Scanning electron microscopy, Fourier-transform infrared spectroscopy, and photoluminescence were used to characterize the resulting structures. The two-dimensional porosity varied only moderately, whereas porous-layer thickness increased with etching time and current density. The silicon oxide to silicon hydride ratio showed comparatively small variations. Corrected photoluminescence spectra were fitted with a four-component quantum-wire/quantum-dot model including localized surface-state transitions. The extracted quantum-wire and quantum-dot dimensions decreased systematically with increasing current density, while the photoluminescence maximum under 375 nm excitation showed no clear monotonic dependence on processing conditions. The results demonstrate that no single descriptor adequately captures the multiscale behavior of pSi.
\end{abstract}

\section{Introduction}
Porous silicon (pSi) is a nanostructured form of crystalline silicon that has attracted sustained interest since the observation of intense visible photoluminescence from electrochemically etched silicon in the early 1990s \cite{Canham1990}. Its distinctive optical properties are associated with the formation of a nanocrystalline silicon skeleton within a porous network, produced by the partial electrochemical dissolution of the crystalline substrate \cite{Canham1990, Cullis1991, Cullis1997, Bisi2000}. The resulting material combines the electronic properties of silicon with a high internal surface area and nanoscale structural features, making pSi attractive for optoelectronics, sensing, photonics, and related applications \cite{Bisi2000}. Because its functional properties depend strongly on pore geometry and porous-layer architecture, electrochemical processing conditions play a central role in determining the final morphology and optical response.

The porous structure is determined by several interdependent parameters of the electrochemical etching process, including current density, electrolyte composition, and etching time. These parameters influence the rate of silicon dissolution and consequently the thickness, porosity, and morphology of the resulting porous layer. Experimental studies on p-type pSi have demonstrated that changes in current density, etching time, and electrolyte composition can produce significant variations in porous-layer thickness and porosity \cite{Wolter2017}. More recent studies likewise emphasize the sensitivity of porous-silicon morphology and optical response to etching conditions \cite{Levy1994, Wang2010, Xu2019, Volovlikova2020, Kuntyi2022, Ptashchenko2021}. Thus, a description based solely on a single morphological parameter such as porosity is insufficient to fully characterize the nanostructured material, particularly when different fabrication conditions can produce comparable apparent porosities while modifying the thickness and nanoscale structure of the porous layer.

In addition to its morphology, the surface chemistry of pSi plays a fundamental role in determining its physical and optical behavior. The high internal surface area of pSi leaves a large fraction of silicon atoms at or near the surface, where they can be terminated by hydrogen immediately after electrochemical etching and subsequently undergo oxidation upon exposure to the environment \cite{Bisi2000}. Fourier-transform infrared (FTIR) spectroscopy provides a direct means of monitoring vibrational signatures associated with silicon hydride (Si–H) and silicon oxide (Si–O) bonds and, consequently, of assessing changes in the chemical termination of the porous structure. Recent experimental work has also linked fabrication-dependent morphological changes with infrared signatures and oxidation behavior in pSi \cite{Xu2019, Volovlikova2020}. The relationship between surface oxidation and optical properties is particularly relevant because oxidation can modify the electronic states associated with nanocrystal surfaces. Experimental studies have demonstrated correlations between surface species, including the relative Si–O and Si–H contributions, and the photoluminescence of pSi \cite{Tsybeskov1994, Rigakis1997}. Recent work performed by our group has further examined spatially resolved oxidation and photooxidation processes in nanocrystalline silicon, providing additional evidence of the importance of surface chemistry in these materials \cite{Ramirez2024}.

Photoluminescence (PL) in pSi is generally associated with radiative recombination processes involving the nanocrystalline silicon skeleton. The quantum-confinement interpretation established in early studies relates the observed visible emission to the nanoscale dimensions of the remaining crystalline silicon structures \cite{Canham1990, Cullis1991, Cullis1997}. Subsequent investigations have shown that pSi can contain nanostructures with different geometries and characteristic dimensions, including structures that can be described in terms of quantum wires (QWs) and quantum dots (QDs). Recent studies continue to demonstrate that electrochemical processing conditions can modify both nanostructural characteristics and PL \cite{Volovlikova2020, Kuntyi2022}. Consequently, the PL spectrum may contain contributions from distinct populations of nanostructures as well as from electronic states associated with their surfaces \cite{Cullis1997, Bisi2000}.

Despite the extensive literature on pSi, establishing direct relationships among fabrication conditions, macroscopic morphology, surface chemistry, nanostructure dimensions, and PL response remains challenging. In particular, variations in PL intensity or spectral position cannot necessarily be attributed to a single parameter such as porosity or oxidation. The different quantities considered here describe different structural scales of the material: scanning electron microscopy (SEM) provides information about the morphology and thickness of the porous layer, FTIR probes the chemical termination of its internal surface, and PL provides information about the electronic and optical properties of the nanoscale silicon structures. Recent reviews emphasize that pSi properties emerge from the coupled effects of processing conditions, pore geometry, layer architecture, and surface chemistry \cite{Kuntyi2022}. A combined analysis of these techniques is therefore necessary to determine whether systematic relationships exist among these different levels of description.

Recent developments in the interpretation of nanocrystalline-silicon PL provide an additional framework for addressing this problem. Ref. \cite{Ramirez2025} proposes an improved model in which the PL spectrum is represented by four contributions: band-to-band recombination in QWs and QDs and localized-to-band recombination associated with surface states in the corresponding nanostructures. The model explicitly incorporates oxide-related surface states and allows characteristic QW and QD dimensions to be extracted from the experimental PL spectra. This approach provides a means of connecting spectroscopic information with both nanostructure dimensions and surface chemistry and is therefore particularly suitable for a correlation-based analysis of pSi \cite{Ramirez2025}.

The present work investigates a systematic set of twenty pSi samples prepared under different electrochemical conditions, with variations in current density, etching time, and electrolyte composition. The samples were characterized by SEM, FTIR spectroscopy, and photoluminescence. SEM measurements were used to determine the two-dimensional surface porosity and porous-layer thickness, while FTIR spectra were analyzed to quantify the relative contributions of Si-O and Si-H through the SiO/SiH ratio. The PL spectra were analyzed using the four-component model of Ref. \cite{Ramirez2025} to obtain characteristic QW and QD dimensions.

The main objective is not simply to characterize the individual properties of the samples, but to investigate possible correlations among the fabrication parameters, porous-layer morphology, surface chemical composition, nanostructure dimensions, and optical response. By combining measurements obtained at different structural scales, this study seeks to determine which properties exhibit systematic relationships and which remain comparatively insensitive to the fabrication conditions investigated. In particular, the analysis examines whether changes in current density, etching time, and electrolyte composition are reflected consistently in the morphology, SiO/SiH ratio, PL response, and characteristic QW/QD dimensions. Such a correlation-based approach is intended to provide a more comprehensive picture of the factors governing the optical behavior of electrochemically produced pSi and to distinguish macroscopic morphological effects from nanoscale and surface-chemical contributions.

\section{Experimental Details}
Samples of pSi were fabricated by electrochemical etching of commercially available boron-doped crystalline silicon wafers with (100) crystallographic orientation and a resistivity ranging from 30 to 50 $\Omega$·cm. Square-shaped specimens with an area of approximately 1 cm² were cut from a 10 cm diameter wafer and employed as starting substrates. The sample production procedure was similar to that previously described in Ref. \cite{Ramirez2024}.

A total of twenty pSi samples were produced by systematically varying three electrochemical parameters: current density, etching time, and acid concentration. Two current densities were employed, namely 23.1 mA/cm$^-2$ (low J) and 53.8 mA/cm$^-2$ (high J). The electrochemical process was performed under galvanostatic conditions using an Agilent 3645A power source. Five etching times were investigated: 7, 10, 20, 30, and 40 min. Two acid concentrations, 12.5\% (in the proportions: [HF:Ethanol:H2O]=[4:11:17]) and 20.0\% ([HF:Ethanol:H2O]=[4:5:11]), were employed. The resulting 2 × 5 × 2 experimental matrix comprised twenty samples, as shown in Table I. The sample labels (AX, BX, CX, DX, with X={0, 1, 2, 3, 4}, X accounting for different etching times) are retained throughout the analysis.

\begin{table}[h]
\centering
\caption{Produced samples for the different parameters employed. For the sample labels, “0” corresponds to 7 min, “1” to 10 min and so forth.}
\begin{tabular}{ | m{3.25cm} | m{4.25cm}| m{4.25cm} | }
\toprule
Acid concentration & Low J: 23.1 mA/cm$^2$ & High J: 53.8 mA/cm$^2$ \\
\midrule
12.5\% & A0, A1, A2, A3, A4: & B0, B1, B2, B3, B4: \\
       & T = 7, 10, 20, 30, 40 min & T = 7, 10, 20, 30, 40 min \\
\midrule
20.0\% & C0, C1, C2, C3, C4: & D0, D1, D2, D3, D4: \\
       & T = 7, 10, 20, 30, 40 min & T = 7, 10, 20, 30, 40 min \\
\bottomrule
\end{tabular}
\end{table}

SEM images were acquired using a JEOL JSM-IT1500 electron microscope. Surface images were used to determine the two-dimensional porosity, while cross-sectional images were used to determine the porous-layer thickness.

FTIR measurements were performed using a PerkinElmer Spotlight 400 instrument equipped with a liquid-nitrogen-cooled mercury cadmium telluride (MCT) detector. The obtained spectra were analyzed by peak deconvolution using Fityk software. The SiO/SiH ratio was obtained by comparing the total integrated area associated with silicon oxide peaks with the corresponding total area of silicon hydride peaks.

PL measurements were carried out with a PerkinElmer PL8500 spectrofluorometer. A variable-angle holder for solid samples was employed. The angle between the excitation beam and the normal direction of the sample was set to 60°, whereas the angle to the collecting port was 45°. The excitation and collecting ports of the instrument are separated by 90°. At a 45° sample-holder configuration, the specularly reflected excitation radiation would enter the collecting port. The sample was therefore rotated by an additional 15°, giving a total angle of 60° with respect to the sample normal, in order to displace the reflected excitation radiation from the collection path and favor detection of the scattered PL. Aperture slits were set to provide a spectral resolution of 5–10 nm, and a scanning speed of 1200 nm/min was used. To eliminate the contribution from scattered excitation light, the response of a crystalline silicon substrate was first recorded under identical experimental conditions. This spectrum was subsequently subtracted from the corresponding pSi measurements, yielding corrected PL spectra representative of the luminescent emission of the porous structures. The corrected PL spectra were analyzed using the model proposed in Ref. \cite{Ramirez2025}, which separates the emission into contributions associated with QW, QD, and localized surface states. The model was used to extract characteristic QW and QD sizes from the experimental spectra.

\section{Results and Discussion}
Representative SEM surface and cross-sectional views of two characteristic samples, B4 and C0, are shown in figure 1.

\begin{figure} 
    \centering
    \includegraphics[width=1.0\linewidth]{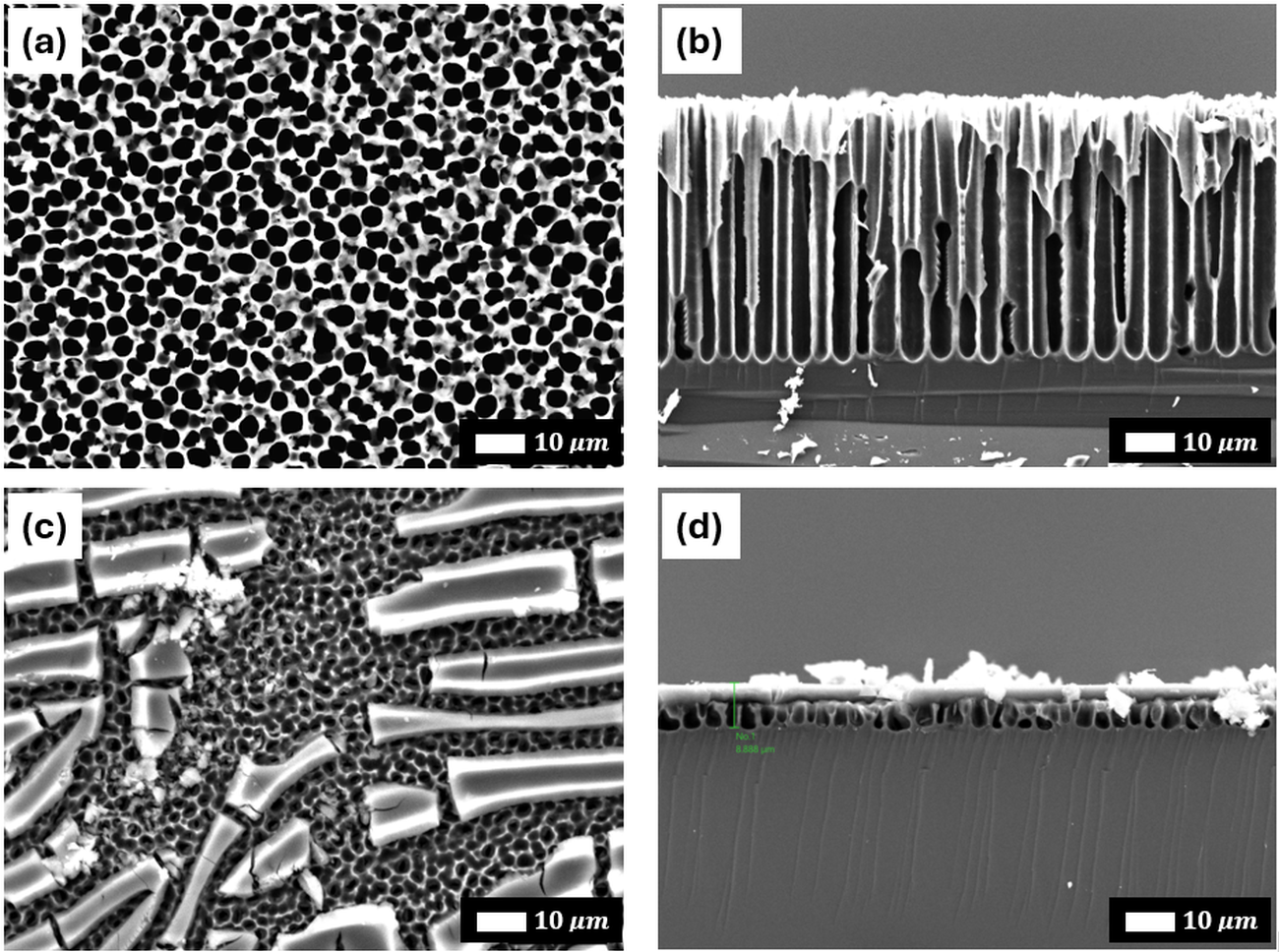}
    \caption{Surface views (upper and lower left) and cross-sectional views (upper and lower right) of the B4 sample (top) and C0 sample (bottom). The horizontal scale bars at the bottom right of the images correspond to 10 $\mu$m.}
    \label{fig:Fig1}
\end{figure}

Some samples exhibit the surface prominences observed in sample C0 (see figure 1(c)), where parts of the surface are covered by silicon residues on top of the porous regions. The cross-sectional image shows a relatively uniform porous layer beneath these features. For such samples, the porosity was calculated using surface regions where the residues were absent. The remaining samples exhibit morphologies like that of B4. The porous layer thickness can be readily determined from the cross-sectional views.

The measurements of two-dimensional porosity and porous-layer thickness for all samples are presented in figure 2 for HF concentrations of 12.5\% and 20.0\%. Because porosity is bounded between 0 and 1 whereas thickness can reach values approaching 100 $\mu$m, logarithmic vertical axes were used to facilitate the simultaneous representation of both quantities. The measured porosity values are relatively similar among the samples, remaining approximately between 0.3 and 0.6. In contrast, the thickness tends to increase with increasing etching time. Higher current density is also associated with greater porous-layer thickness than lower current density.

\begin{figure} 
    \centering
    \includegraphics[width=1.0\linewidth]{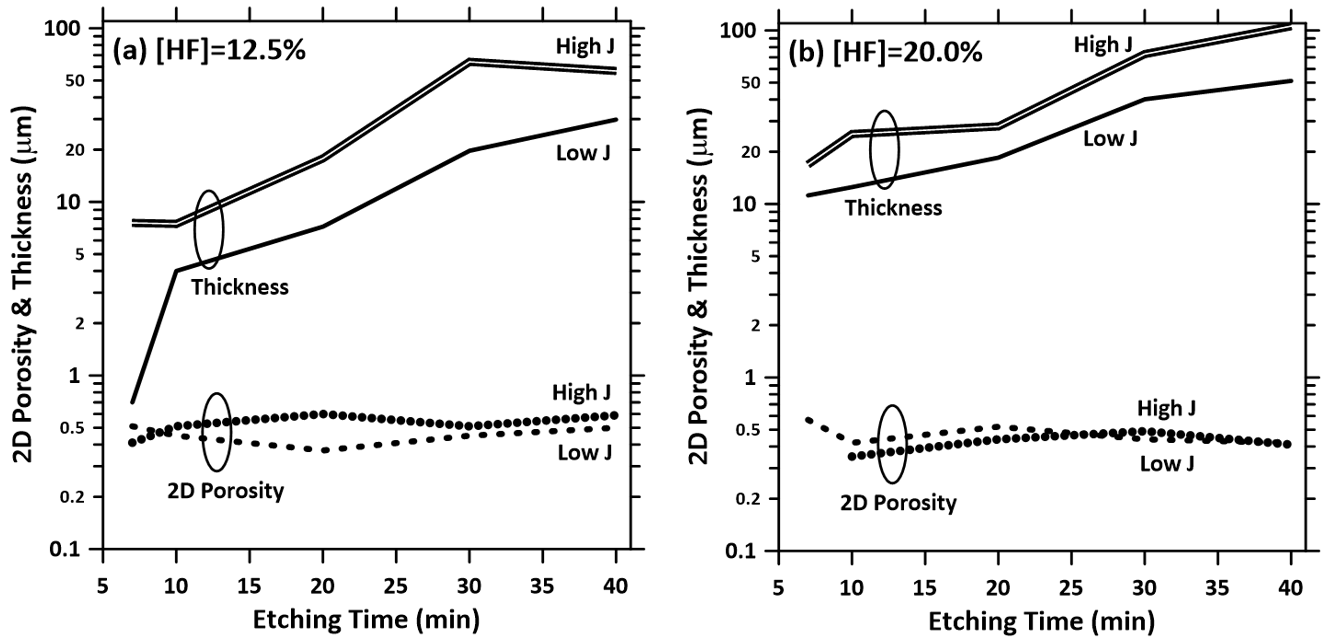}
    \caption{Plot of the measurements of 2D Porosity and Thickness for all the samples in the case of HF concentration of 12.5\% (a) and 20.0\% (b). Dotted lines correspond to porosity and full lines to thickness for both cases of low and high current densities, as marked.}
    \label{fig:Fig2}
\end{figure}

An FTIR absorption spectrum of sample A1 is shown in figure 3. All samples exhibited silicon hydride and silicon oxide stretching bands in the region from approximately 2400 to 2000 cm$^{-1}$, together with a silicon oxide stretching band in the 1160–1050 cm$^{-1}$ region, consistent with the assignments reported for pSi \cite{Bisi2000, Xu2019, Volovlikova2020}. For sample A1, the oxide- and hydride-related contributions in the 2350–2000 cm$^{-1}$ range were deconvoluted into Gaussian peaks using Fityk. The SiO/SiH ratio was calculated by comparing the total integrated area under the silicon oxide peaks with the total integrated area under the silicon hydride peaks.

\begin{figure} 
    \centering
    \includegraphics[width=1.0\linewidth]{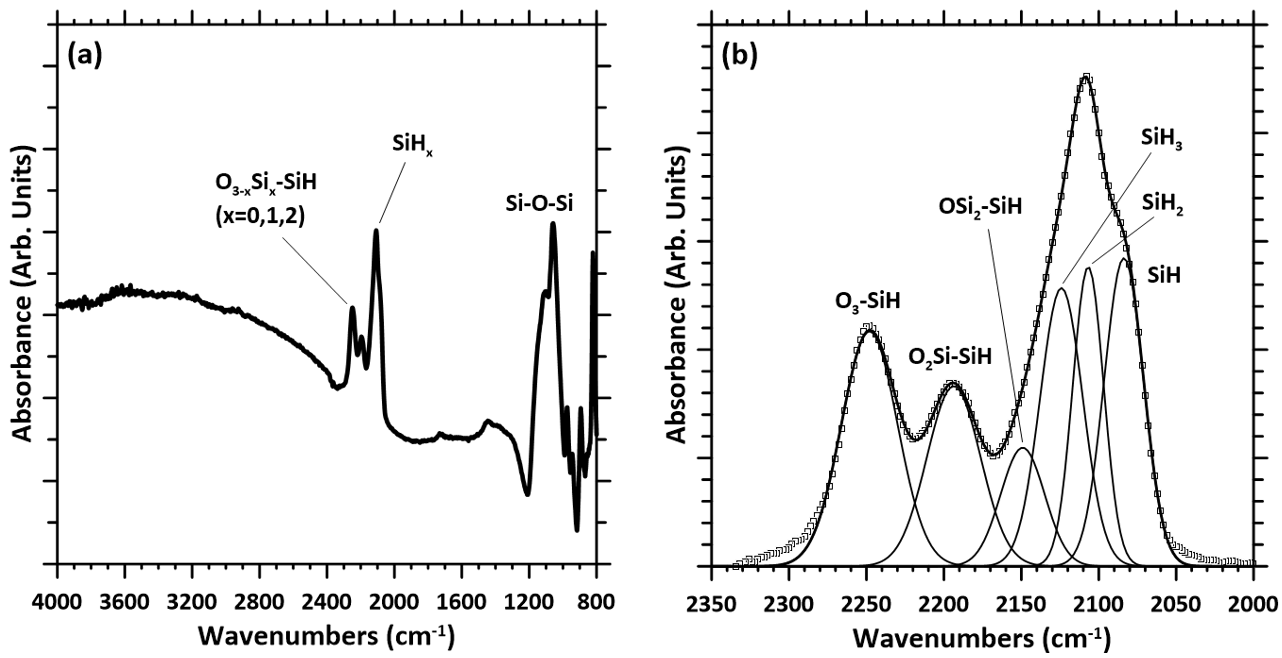}
    \caption{FTIR absorption spectrum of the A1 sample in the range of 4000 to 800 cm$^{-1}$ (a), and in the 2350 to 2000 cm$^{-1}$ range (b), corresponding to the silicon oxide and silicon hydride bands. In (b), the fitted Gaussian peaks corresponding to different oxide- and hydride-related contributions are marked on the plot.}
    \label{fig:Fig3}
\end{figure}

The PL spectra of sample B1 for excitation wavelengths of 275, 375, and 425 nm are presented in figure 4(a). A slight red shift of the emission maximum is observed as the excitation wavelength increases from 275 to 425 nm. For the 375 nm excitation spectrum, figure 4(b) shows the deconvolution into four components according to the model proposed in Ref. \cite{Ramirez2025}. The four contributions are associated with QW and QD transitions and localized states at the corresponding QW and QD surfaces. The calculated fit agrees well with the experimental data. The use of this model provides a direct link between the measured PL spectrum and characteristic nanostructure dimensions, complementing the morphological information obtained from SEM \cite{Volovlikova2020, Ramirez2025}.

\begin{figure} 
    \centering
    \includegraphics[width=1.0\linewidth]{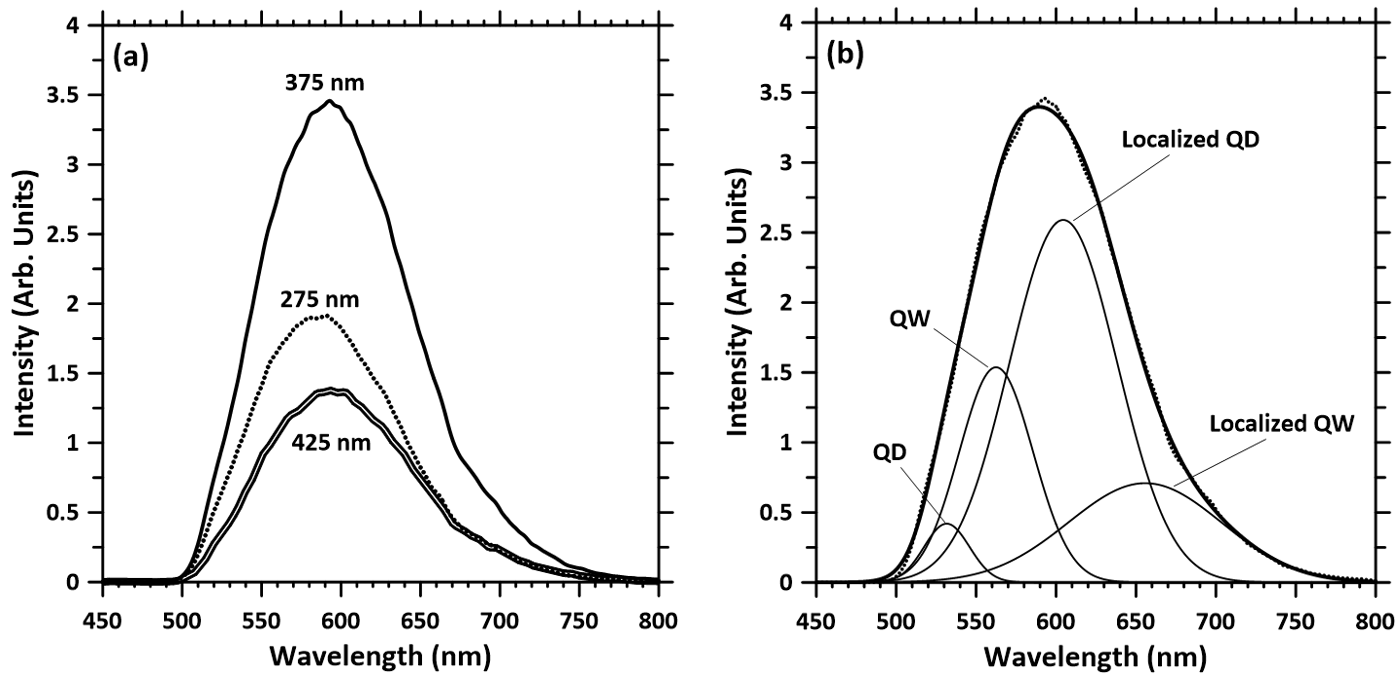}
    \caption{PL spectra of sample B1 for three excitation wavelengths: 275 nm (dotted plot), 375 nm (line plot), and 425 nm (double-line plot) as marked in (a), and the PL spectrum for 375 nm excitation in (b). The four components associated with QW and QD, as well as with localized states at the QW and QD surfaces, are marked in the figure.}
    \label{fig:Fig4}
\end{figure}

Figure 5 compares the SiO/SiH ratio and normalized PL maximum for all samples under 375 nm excitation, for HF concentrations of 12.5\% and 20.0\%. A small increasing tendency in the SiO/SiH ratio is noticeable for both groups of samples, although the relative behavior of the low- and high-current-density series changes between the two HF concentrations. Overall, the oxide bands maintain a broadly similar proportion relative to the hydride bands, indicating that no pronounced increase in relative oxidation occurs with increasing etching time. On the other hand, the PL maximum under 375 nm excitation shows no clear monotonic tendency with etching time or current density.

\begin{figure} 
    \centering
    \includegraphics[width=1.0\linewidth]{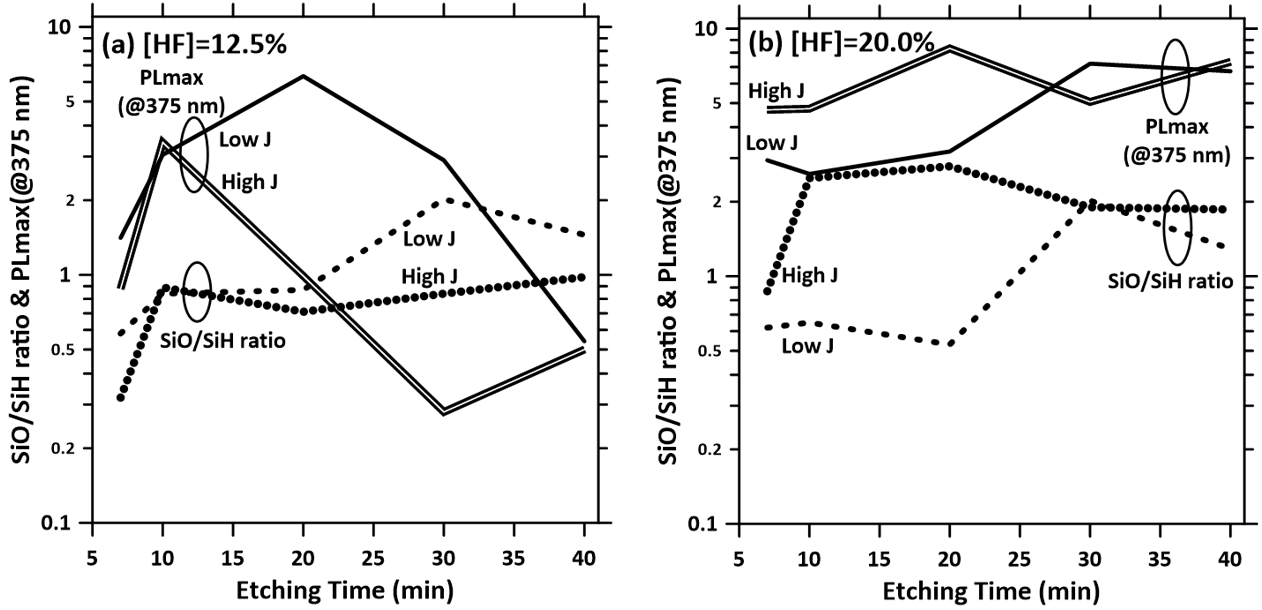}
    \caption{SiO/SiH ratio and normalized PL maxima for all samples under 375 nm excitation in the case of HF concentrations of 12.5\% (a) and 20.0\% (b). Dotted lines correspond to the SiO/SiH ratio and solid lines to the normalized PL maxima for both low- and high-current-density cases, as marked.}
    \label{fig:Fig5}
\end{figure}

The extracted mean QW and QD sizes for all samples are shown in figure 6. Overall, QW tend to have larger diameters than QD. A second trend is that samples produced at higher current densities exhibit smaller characteristic diameters than those produced at lower current densities. This behavior is observed for both HF concentrations.

\begin{figure} 
    \centering
    \includegraphics[width=1.0\linewidth]{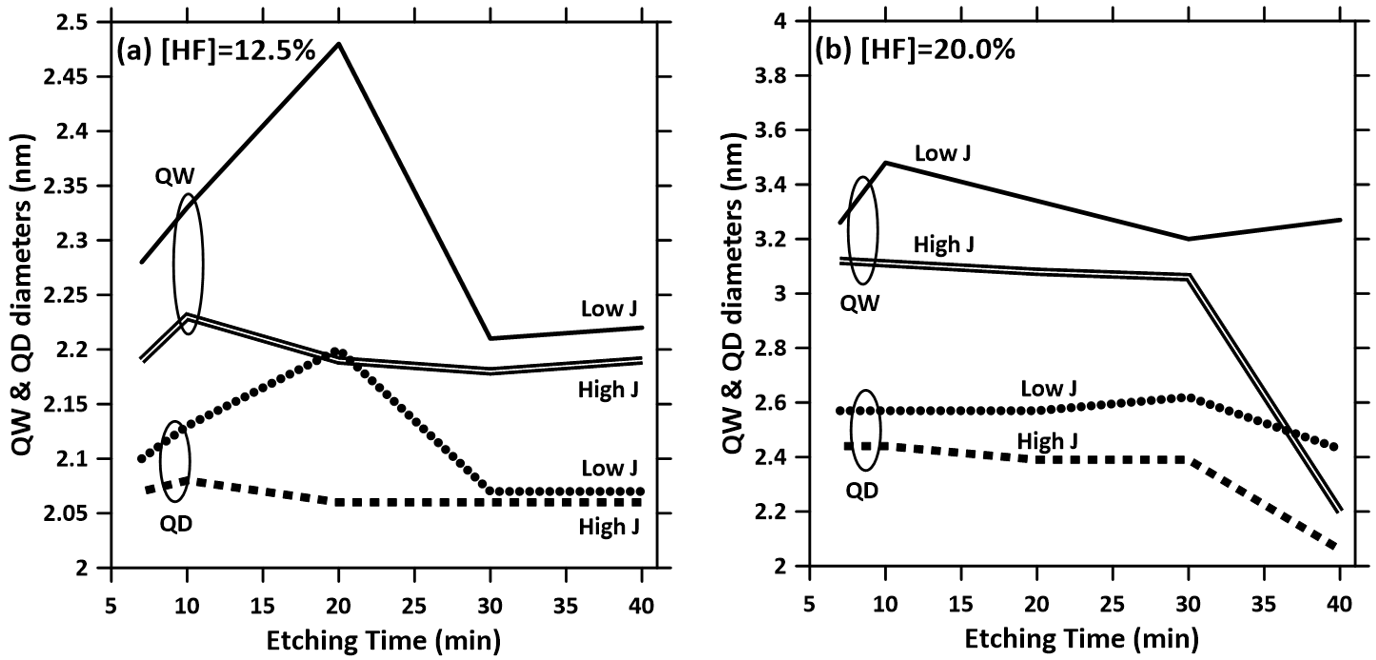}
    \caption{Extracted mean nanocrystal QW and QD sizes for all samples in the case of HF concentrations of 12.5\% (a) and 20.0\% (b). Dotted lines correspond to QD and solid lines to QW for both low- and high-current-density cases, as marked.}
    \label{fig:Fig6}
\end{figure}

Taken together, the SEM, FTIR, and PL measurements indicate that the fabrication parameters do not affect all measured properties in the same way. The porous layer thickness shows a clear dependence on etching time and current density, whereas the measured two-dimensional porosity remains within a comparatively limited range. This distinction is consistent with the established sensitivity of porous layer architecture to electrochemical processing conditions \cite{Wolter2017, Xu2019, Kuntyi2022}. The FTIR measurements indicate only modest variation in the relative oxide and hydride contributions. In contrast, the PL-derived characteristic QW and QD dimensions display a systematic dependence on current density, while the PL maximum itself does not exhibit a simple monotonic dependence on the fabrication parameters.

These observations suggest that the electrochemical parameters influence different structural scales of pSi rather than producing a single correlated response. The layer thickness reflects the extent of electrochemical dissolution, while the characteristic dimensions extracted from the PL model provide information about the nanoscale silicon structures. At the same time, the relatively limited variation of the SiO/SiH ratio indicates that changes in the PL response cannot be attributed solely to a progressive increase in oxidation. The combined measurements therefore provide a more complete description of the material than any single characterization technique alone, in agreement with the multiscale view of pSi formation emphasized in recent literature \cite{Volovlikova2020, Kuntyi2022}.

An important feature of the present results is that comparable two-dimensional porosities can coexist with different porous-layer thicknesses and different characteristic QW/QD dimensions. Likewise, samples prepared under different electrochemical conditions may display similar normalized PL maxima. This lack of a simple one-to-one relationship among the measured quantities suggests that morphology, surface chemistry, and nanostructure dimensions contribute simultaneously to the optical response. These observations also caution against interpreting the PL maximum alone as a direct proxy for either porosity or surface oxidation.

\section{Conclusions}
Twenty pSi samples were prepared by electrochemical etching of boron-doped (100) crystalline silicon under two current densities, two HF concentrations, and five etching times. SEM, FTIR, and PL measurements were combined to investigate possible correlations among electrochemical processing, porous-layer morphology, surface chemistry, and nanoscale optical properties.
SEM analysis showed that the two-dimensional porosity remained within approximately 0.3–-0.6 for the investigated samples, whereas porous layer thickness tended to increase with etching time and was higher at the larger current density. Thus, within the experimental range studied, layer thickness responds more clearly to the electrochemical conditions than the measured two-dimensional porosity.

FTIR measurements confirmed the presence of both Si–H- and Si–O-related vibrational contributions. The SiO/SiH ratio exhibited only modest changes across the sample set, with no pronounced increase in relative oxidation as etching time increased. Consistently, the normalized PL maximum under 375 nm excitation did not display a clear monotonic dependence on etching time or current density.

The four-component PL model provided characteristic QW and QD dimensions for the samples. QW dimensions were generally larger than QD dimensions, while higher current density was associated with smaller extracted characteristic dimensions. This indicates that PL analysis provides nanoscale structural information that is not directly represented by the measured two-dimensional porosity.

Overall, the results support a multiscale interpretation of pSi in which electrochemical processing, morphology, surface chemistry, nanostructure dimensions, and PL are related but are not governed by a single controlling parameter. The combined analysis of these properties across a systematic twenty sample matrix provides evidence that correlations exist at different structural levels, while also showing that some commonly considered descriptors, such as two-dimensional porosity or the PL maximum, do not by themselves capture the full response of the material. Further quantitative analysis will be required to establish the specific mechanisms connecting these properties.

\section*{Acknowledgments}
This work was partially supported by the Vicerrectoría de Investigación of the Universidad de Costa Rica under project No. C6-076.

\bibliographystyle{unsrt}
\bibliography{References}

\end{document}